\documentstyle[12pt,aaspp4]{article}

\lefthead{M.P.Allen \& J.E.Horvath}
\righthead{Implications of a constant braking index}

\begin{document}
\title {Implications of a constant observed braking index \\
for young pulsar's spindown}
\author{M.P.Allen\altaffilmark{1} and J.E.Horvath\altaffilmark{2}}
\affil{Instituto Astron\^omico e Geof\'\i sico \\
Universidade de S\~ao Paulo - Av. M.St\'efano 4200 - \'Agua Funda \\
(04301-904) S\~ao Paulo SP - Brasil}
\altaffiltext{1}{mpallen@iagusp.usp.br}
\altaffiltext{2}{foton@iagusp.usp.br}

\begin{abstract}

The observed braking index $n_{obs}$ which had been determined for a few
young pulsars, had been found to differ from the expected value for a 
rotating magnetic dipole model. In addition, the observational jerk
parameter,
determined for two of these pulsars, disagrees with the theoretical
prediction 
$m_{obs}$ = 15 in both cases. We propose a simple model able to account
for 
these differences, based on a growth of the "torque function" $K \, = \,
- \dot
\Omega / \Omega^{n}$, under the constraint that $n_{obs}$ is 
a constant. We show that there is observational evidence supporting the
latter 
hypotesis, and derive initial values for several physical quantities
for the four pulsars whose $n_{obs}$ have been measured. 

\end{abstract}


\section{Introduction}

The well-known vacuum dipole model expresses the external torque acting
on 
pulsars as

$$I \, \dot \Omega \, = \, - {2 \over {3 \, c^{3}}} \, B^{2} R^{6}
\sin^{2} \alpha \, \Omega^{n} \eqno(1)$$
\noindent
where $\alpha$ is the angle between $B$ and $\Omega$, $B$ and $R$ are 
the magnetic field and the radius of the star respectively, $I$ is the
star's moment of inertia and $c$ is the velocity of light. The braking 
index $n$ is usually assumed to be 3 as predicted by the magnetic dipole
model 
(see e.g. Manchester \& Taylor 1977), but actually the four (young) 
pulsars with the best determinations of braking index show values $<3$ 
(see Blandford \& Romani 1988 and references therein). In the general 
literature all these quantities are taken to be constants (except of
course 
for $\Omega$ and its derivatives), which allows one to rewrite eq.(1) as

$$\dot \Omega \, = \, -K \, \Omega^{n} \eqno(2)$$   
\noindent
where $K$ is a function that absorbs the structural factors, and we will 
refer to throughout this work as the "torque function". 

There is evidence that in 1975 and 1989 the spin rate $\Omega$ of the
Crab 
pulsar suddenly increased by amounts $\Delta \Omega / \Omega \, 
\sim \, 10^{-8}$ and after that continued to spin-down at a faster rate 
(that is, the pulsar continued to spin {\it slower} than before) of 
$\Delta \dot \Omega / \dot \Omega \, \sim \, 10^{-4}$ (Gullahorn et al. 
1977, Lohsen 1981, Lyne, Smith \& Pritchard 1992), characterizing a 
{\it permanent} deficit in $\dot \Omega$. The same feature is also
present 
in the 1969, 1981 and 1986 glitches (Lyne \& Pritchard 1987, Lyne, 
Pritchard \& Smith 1993). From eq.(1) it is clear that the peculiar
events 
of 1975 and 1989 require either a reduction of $I$ (Alpar \& Pines
1993),
an increase of the magnitude of the magnetic field $B$ (Blandford, 
Applegate \& Hernquist 1983, Muslimov \& Page 1996, Camilo 1997) or an 
increase of the angle $\alpha$ (Macy 1974). Other dynamical models in
which 
the angle between the magnetic dipole and the rotation axis of the
pulsar 
is allowed to vary hace been proposed by Link \& Epstein (1997) and
Allen 
\& Horvath (1997). In our previous work, we assumed several laws for the 
growth of the angle $\alpha$ and verified which ones provided 
consistent solutions. It was shown that a exponential angular growth
with 
a $e$-folding time of $\sim 10^{4} \, yr$ could fit the small braking 
indexes and jerk parameters of the Crab, Vela and B0540-69 pulsars, 
and although PSR B1509-58 was best-fitted by a logarithmic law, an 
exponential one is not ruled out because of the large uncertainties (25
\%) 
on its jerk value.

The key feature here is the variation of the torque function $K$, which 
cannot be held constant, but instead evolves, growing on timescales 
shorter than the pulsar's life span. This work presents a simple general 
model for torque function growth, based on the observational feature
$n_{obs} 
\, = \, constant$ (see also Ruderman 1993 and references therein). In
the next 
Sections we will show that this simple hypothesis is enough to determine
the 
general behaviour of the pulsar dynamics, along with its evolution on
the 
$\dot P \times P$ diagram, yet taking into account the low braking
indexes 
displayed by young pulsars.

\section{Consequences of $n_{obs}$ = constant}

\subsection{General model}

To relate the model to the observations, we need to define some 
physical quantities. The first one is the variation of the torque
function 
$\Delta K$ due to persistent shifts $\Delta {\dot \Omega}/ {\dot
\Omega}$ 
and $\Delta \Omega / \Omega$, which is easily found from eq.(2) to be

$$\Delta K \, = \, {\biggl( {{\Delta {\dot \Omega}} \over {\dot \Omega}}
- 3 
{{\Delta {\Omega}} \over {\Omega}} \biggr)} \, K . \eqno(3)$$

Dividing eq.(3) by $\Delta t$, here defined as a typical timescale 
between glitches, we obtain a mean increase rate which we shall denote 
as $\bigl< {{\Delta K} \over {\Delta t}} \bigr>$. Note that even the 
exact definition of the mean is not completely satisfactory given the 
different coverages and observational biases for each pulsar (we will 
return to this point below). Using the data from Lyne, Pritchard \& 
Smith (1993) (hereafter LPS), it can be found that the mean torque
function 
variation caused by glitches ({\it discrete} variation) for the Crab is

$${\biggl< {{\Delta K} \over {\Delta t}} {1 \over K} \biggr>}
\, \simeq \, 3 \times 10^{-5} \, yr^{-1} \eqno(4)$$
\noindent
with $\Delta t \, = \, 4.6 \, yr$. 

It can be shown that even if the power of $\Omega$ in the torque 
expression is exactly 3, the observed braking index defined as 
$n_{obs} \, = \, \ddot \Omega \, \Omega / \dot \Omega^{2}$ is 

$$n_{obs} \, = \, 3 \, + \, {\Omega \over {\dot \Omega}} \, {{\dot K}
\over K} 
\, , \eqno(5) $$
\noindent
and the observed jerk parameter defined as $m_{obs}$ =
\hbox{${\Omega}\!\!\!\!^{^{^{...}}}$}$\Omega^{2} / {\dot \Omega}^{3}$ is

$$m_{obs} \, = \, 3 {\bigl( 3 n_{obs} -4 \bigr)} \, + \, {\bigl( n_{obs}
-3 
\bigr)}^{2} \, {\biggl( {{\ddot K \, K} \over {\dot K^{2}}} \biggr)} . 
\eqno(6)$$

Eq.(5) can be inverted to find the ({\it continuous}) torque function 
variation needed to account for the observed braking index of Crab

$${{\dot K} \over K} \, = \, (n_{obs} -3) \, {{\dot \Omega} \over
\Omega} \, 
\simeq \, 1.9 \times 10^{-4} \, yr^{-1} . \eqno(7)$$

So, the discrete contribution is only $\sim$ 15\% of the continuous 
torque function variation. That means that even pulsars that have not
shown 
any glitches could have their torque functions varying in a continuous
way, and 
also satisfy $n_{obs} <3$. In fact, PSR B1509-58 and PSR B0540-69 have 
never displayed glitches, yet their braking indexes are 2.8 and 2.0 
respectively, although the hypothesis that they are actually glitching
at a
rate comparable to the Crab is not ruled out. Application of eq.(7) to
Vela, 
PSR B1509-58 and PSR B0540-69 yields $6.98 \times 10^{-5}$, $5.25 \times 
10^{-5}$ and $28.8 \times 10^{-5} \, yr^{-1}$, respectively.
 
With the expression for $n_{obs}$ we derive a form for the jump of the 
latter in each glitch event, namely           

$$ \Delta n_{obs} \, = \, n_{obs} \, {\biggl( {{\Delta \Omega} \over
\Omega} \,
-2 \, {{\Delta \dot \Omega} \over {\dot \Omega}} \, + \, {{\Delta \ddot
\Omega}
 \over {\ddot \Omega}} \biggr)} \, . \eqno(8) $$

The division of eq.(8) by $\Delta t$ yields a mean variation rate of
$n_{obs}$,
which can be calculated from LPS for the Crab to be

$${\biggl< {{\Delta n_{obs}} \over {\Delta t}} \biggr>} \, \simeq \,
-1.5 \times
10^{-4} \, yr^{-1} \eqno(9)$$
\noindent
assuming that $\ddot \Omega$ did not vary in the events, as assumed in
all 
data analysis performed until now. According to LPS the data show that 
$n_{obs}$ appears to be constant within 0.5\% ($\sim 0.01$) over 20
years, 
implying $\left\vert \dot n_{obs} \right\vert \, < \, 5 \times 10^{-4}
\, 
yr^{-1}$, which is consistent with eq.(9). Thus we are led to explore a
scenario in which pulsars evolve along a constant braking index value
$\not =$
3.

A direct measurement of $\dot n_{obs}$ in term of observables can be
obtained as

$$\dot n_{obs} \, = \, n_{obs} \, {\biggl(
{{\hbox{${\Omega}\!\!\!\!^{^{^{...}}}$}} \over {\ddot \Omega}} \, + \, 
{{\dot \Omega} \over \Omega} \, -2 \, {{\ddot \Omega} \over {\dot
\Omega}} 
\biggr)} \, = \, {{m_{obs} -n_{obs}(2n_{obs}-1)} \over {\Omega / \dot 
\Omega}}. \eqno(10)$$

Setting $\dot n_{obs}$ = 0, we can express $m_{obs}$ as function of
$n_{obs}$, 
$\dot \Omega$ and $\Omega$, and replace it in eq.(6), yielding

$${{\ddot K \, K} \over {\dot K^{2}}} \, = \, 1 \, - \, 
{{n_{obs} -1} \over {3-n_{obs}}} \, = \, constant \eqno(11)$$
\noindent
which can be now integrated from an arbitrary point to the present
values
(indicated by the subscript $p$), resulting in

$$\dot K \, = \, \dot K_{p} {\biggl( {K \over K_{p}} \biggr)}^{1 \, - \, 
{{n_{obs} -1} \over {3-n_{obs}}}} . \eqno(12)$$

Eq.(5) can be also integrated in the same way, and we obtain

$${\biggl( {\Omega \over \Omega_{p}} \biggr)}^{n_{obs} -3} \, = \, {K
\over
K_{p}} . \eqno(13)$$

Again, integration is performed and with the help of eq.(7) the time
dependence 
of $K$ is found to be (for $n_{obs} \, \not =$ 1)

$${K \over K_{p}} \, = \, {\biggl[ 1+(n_{obs} -1)
(t_{p} -t) {{\dot \Omega_{p}} \over \Omega_{p}} \biggr]}^{{{3- n_{obs}}
\over
{n_{obs} -1}}} \eqno(14)$$
\noindent
where $t_{p}$ is the true age of the pulsar; the time dependence 
of $\Omega$ is trivially recovered substituting eq.(14) into eq.(13)

$${\Omega \over \Omega_{p}} \, = \, {\biggl[ 1+(n_{obs} -1) (t_{p} -t)
{{\dot 
\Omega_{p}} \over \Omega_{p}} \biggr]}^{{1 \over {1- n_{obs}}}} .
\eqno(15)$$

It is worth remarking that eq.(15) has the same form of the standard
calculation
made by assuming that the torque function does not change, with 
$n_{obs}$ replacing $n=3$ (see Manchester \& Taylor 1977). In other
words, 
it appears as though the actual external torque acting on pulsars has
the 
form $\dot \Omega \, \propto \, \Omega^{n_{obs}}$. Therefore, 
several approximations customarily made in the literature are exact in
the 
limit $n_{obs} \, = \, constant$.

Finally, from eq.(10) we write

$$m_{obs} \, = \, n_{obs} (2 n_{obs} -1) . \eqno(16)$$

Again, we note that eq.(16) has the same form as the conventional
relation,
if we replace $n=3$ by $n_{obs}$.

It is important to note that the lower $n_{obs}$ is, the closer the
initial
period is to the present one. Therefore, the use of the characteristic
age 
$\tau \, = \, {1 \over {(n_{obs} \, - \,1)}} \, {\Omega \over
{\left\vert 
\dot\Omega \right\vert}}$ (obtained from eq.(15)), though strictly valid
when
$n_{obs}$ = constant, introduces a non-negligible error because $\Omega
\,
\simeq \, \Omega_{o}$.

The important fact is that postulating $\dot n_{obs}$ = 0 (as suggested
by
observations) completely determines the dynamical evolution of the
pulsar,
whichever model of variation one chooses between magnetic field, moment
of
inertia or angle $\alpha$. In Table 1 we show the correspondence 
between $K$ and its derivatives for all specific models.

In Fig.1 we depict the evolutionary tracks of the Crab and B1509-58 
pulsars predicted by our model, compared to the standard model tracks, 
from birth to $5 \times 10^{4}$ years. Fig.2 shows the same for Vela 
and B0540-69. Remarkably, $\dot P$ for Vela is actually increasing with
time, 
although this does not mean a rising torque, because of the 
compensating effect of the power of $\Omega$.  

Now, it is easy to calculate the initial period for any pulsar whose
$n_{obs}$ 
has already been determined. In Table 2 we show observed and calculated
quantities for the above pulsars,
where $L\, = \, I \Omega \dot \Omega$ is the energy-loss rate of the
rotating 
magnetic dipole. The data used in calculations came from LPS (Crab),
Kaspi et 
al.(1994) (PSR B1509-58), Taylor et al. (1995, unpublished, see Taylor, 
Manchester \& Lyne 1993) (PSR B0540-69), Lyne et al. (1996) and Taylor, 
Manchester \& Lyne (1993) (Vela). Except in the case of Crab, the
conventional 
characteristic age $\Omega / (2 \dot \Omega)$ was arbitrarily used as
the true 
age of the pulsar, lacking a more reliable determination.

\subsection{Crab Pulsar (PSR B0531+21)}

The Crab pulsar is known to be among the most active glitching pulsars. 
If the glitch activity has been aproximately the same since its 
birth, $n_{obs}$ would have decreased by an amount ${\bigl< \Delta
n_{obs} 
/ \Delta t \bigl>} \times t_{p} \, \sim \, 0.15$, or 6\% of the present 
value. In the 25 years that have passed since the first Crab
measurements, 
glitches could have provoked a reduction of only 0.004 (0.2\%) in
$n_{obs}$, 
far below the observational limit (see LPS). About 20 years' more data
will 
be needed to unambigously detect $n_{obs}$ variation.
   
The values obtained assuming $n_{obs} \, = \, constant$ (Table 2) 
are very close to those observed, which is as expected, 
since \hbox{${\Omega}\!\!\!\!^{^{^{...}}}$} was in fact estimated by LPS
using
eq.(16). The good fit to the data reinforces that $\dot n_{obs}$ is zero 
or very small. 

\subsection{Vela Pulsar (PSR B0833-45)}

Some authors (Aschenbach, Egger \& Tr\"umper 1995, Lyne et al. 1996) 
have discussed the possibility of Vela being up to 3 times older than 
its conventional characteristic age. As a test of the sensitivity of
this
model to the true age adopted, an age doubling alters the values in 
Table 2, as follows

$$P_{o} \, \simeq \, 26 \, ms$$

$$K_{o} \, \simeq \, 1.56 \times 10^{-15} \, s \eqno(17)$$

$$\dot P_{o} \, \simeq \, 60.4 \times 10^{-15}$$

$${L_{o} \over L_{p}} \, \simeq \, 20$$
                                 
It is worth to noting that Vela is also one of the most glitch-active
pulsars 
known, and, like the Crab, there could be a discrete torque function
growth, as
discussed in the introduction, which would make $\dot n_{obs} \, \not =
\, 0$. 
However, timing analyses made to date (Lyne et al. 1996 being the most
recent) 
have failed to unambigously detect {\it permanent} shifts in $\dot
\Omega$, 
because the typical (non-permanent) shift $\Delta \dot \Omega / \dot
\Omega$ 
is $10^{-3}$ and the mean time between glitches is $\sim$ 2 yr, leading
to 
the conclusion that if such a permanent component does exist, it must be 
negligible.

\section{Conclusions}

We have shown that low braking indices, observed in young pulsars, can
be 
attributed to a variation of the torque function $K$. Besides, there are 
evidence that $n_{obs} \, = \, constant$ in some of these pulsars, and
this 
hypothesis leads to a complete determination of their dynamics, which we
have
calculated. Our expressions are almost identical to the conventional
ones, 
when $n$ substitutes $n_{obs}$, justifying the customarily estimates.

From Table 2, we can see that $K$ must have increased by 25\% to 100\%
since 
the pulsar birth. From the equivalence in Table 1, we verify that this
means a 
reduction of 20-50\% in the moment of inertia, which can be accomodated 
neither in the current vortex creep model nor as loss of oblateness. So
a
decrease in the moment of inertia cannot be invoked as the main physical
agent 
behind torque evolution, although it probably is in the case of
glitches.

An inspection of Table 2, Fig.1 and Fig.2 reveals sistematically larger
initial 
periods and smaller period derivatives than those obtained in the
standard 
model. It becomes clear that, within $n_{obs} \, = \, constant$ models,
the 
characteristic age is poorly related to the true age of pulsars below
$10^{5} 
\, yr$. This may affect statistical studies of pulsar populations.

We think this torque evolution, especially if driven by angular growth, 
probably stops at ages of the order of $10^{4}$ years, when $\alpha \, =
\,
90^{o}$ (Allen \& Horvath 1997). Further improvement will require 
observations of the braking indices of middle-aged ($\sim 10^{5}$ years) 
pulsars. 
    
\acknowledgments
We would like to acknowledge J.A. de Freitas Pacheco and an anonymous
referee 
for crytical reviews of this work. 
This work was partially supported by the CNPq (Brazil) through a 
Research Fellowship (J.E.H.), CAPES (Brazil) to M.P.A. and FAPESP
(Brazil).

\clearpage
\begin{deluxetable}{crrr}

\tablecaption{Equivalence between expressions of general and specific
models
\label{tabela1}}

\tablehead{\colhead{General} & \colhead{Angle} & \colhead{Magnetic
Field} & 
\colhead{Moment of Inertia}}
\startdata
K & $\sin^{2} \alpha$ & $B^{2}$ & $I^{-1}$ \nl
$\dot K / K$ & $2 \dot \alpha / \tan \alpha$ & $2 \dot B / B$ & $- \dot
I / 
I$ \nl
$\ddot K / \dot K$ & $2 \cos (2 \alpha) \dot \alpha + (\ddot \alpha /
\dot 
\alpha)$ & $(\dot B / B) + (\ddot B / \dot B)$ & $(-2 \dot I / I) -
(\ddot I / 
\dot I)$ \nl
$\ddot K K / \dot K^{2}$ & $\tan \alpha {\bigl[ \cos (2 \alpha) 
+ (\ddot \alpha / 2 \dot \alpha^{2}) \bigr]}$ & ${1 \over 2} {\bigl( 
1 + (\ddot B B / \dot B^{2}) \bigr)}$ & $2 + (\ddot I I / \dot 
I^{2})$ \nl
\enddata

\end{deluxetable}
\clearpage
\begin{deluxetable}{crrrr}

\tablecaption{Comparison between observations and calculated values for
young
pulsars \label{tabela2}}
\tablehead{\colhead{} & \colhead{Crab} & \colhead{PSR B1509-58} &
\colhead{PSR 
B0540-69} & \colhead{Vela}}
\startdata
$t_{p} \, [yr]$ & 915 & 1550 & 1660 & 11000 \nl
$n_{obs}$ & $2.5179 \pm 0.0001$ & $2.837 \pm 0.001$ & $2.04 \pm 0.02$ & 
$1.4 \pm 0.2$ \nl
$\dot n_{obs} \, [10^{-4} \, yr^{-1}]$ & $-2.7 \pm 1.3$ & $-4 \pm 11$ & 
\nodata & \nodata \nl
$m_{obs}$ & $10.23 \pm 0.03$ & $14.5 \pm 3.6$ & \nodata & \nodata \nl
$m_{obs}$ if $\dot n_{obs} = 0$ & $10.160 \pm 0.001$ & $13.26 \pm 0.01$
& 
$6.28 \pm 0.16$ & $2.8 \pm 0.85$ \nl
$P_{o} \, [ms]$ & 19 & 39 & 25 & 52 \nl
$\dot P_{o} \, [10^{-15}]$ & 559 & 4760 & 492 & 91 \nl
$L_{o} / L_{p} (I = const)$ & $6.7$ & $180$ & $8.4$ & $3.7$ \nl
$K_{o} [10^{-15} \, s]$ & 10.8 & 186 & 12.3 & 4.73 \nl
$K_{o} / K_{p}$ & 0.770 & 0.802 & 0.510 & 0.424 \nl
${{K \ddot K} \over {\dot K^{2}}}$ ($\dot n_{obs} = 0$) & -2.15 & -10.27
& 
-0.081 & 0.737 \nl
\enddata
\end{deluxetable}

\clearpage
\figcaption[ncte1.ps]{In this diagram, the Crab pulsar track is marked
with x,
while PSR B1509-58 is marked with +. Each step is $10^{3} \, yr$, from
birth to
5 $\times 10^{4}$. Squares indicate the present situation. Both tracks
are
intercepted by lines representing the same tracks after the conventional
model.}               

\clearpage
\figcaption[ncte2.ps]{In this diagram, the Vela pulsar track is marked
with x,
while PSR B0540-69 is marked with +. Each step is $10^{3} \, yr$, from
birth to
5 $\times 10^{4}$. Squares indicate the present situation. Both tracks
are
intercepted by lines representing the same tracks after the conventional
model.}           

\end{document}